\documentclass[
    ,final            
  ,mathfrak]
  {aipproc}

\layoutstyle{6x9}

\usepackage{epsfig,amsmath,amssymb}
\usepackage{graphicx}

\newcommand{\ep}{\varepsilon}

\newcommand{\dr}{\delta\omega^{(r)}}

\begin{document}

\title{Multiphoton Antiresonance and
Quantum Activation in Driven Systems }

\author{M. I. Dykman}{
  address={Department of Physics
and Astronomy, Michigan State
University, East Lansing, MI 48824, USA}
}

\begin{abstract}
We show that nonlinear response of a quantum oscillator displays
antiresonant dips and resonant peaks with varying frequency of the
driving field.  The effect is a consequence of special symmetry and is
related to resonant multiphoton mixing of several pairs of oscillator
states at a time. We also discuss escape from a metastable state of
forced vibrations. Two important examples show that the probability of
escape via diffusion over quasienergy is larger than via dynamical
tunneling provided the relaxation rate exceeds both of them.
Diffusion dominates even for zero temperature, so that escape occurs
via quantum rather than thermal activation. The effects can be studied
using Josephson junctions and Josephson-junction based systems.

\end{abstract}

\maketitle


\section{Introduction}
Much progress has been made recently in experimental studies of
periodically modulated vibrational systems. Examples include optically
bistable systems, electrons in Penning traps, Josephson junctions, and
various nano- and micro-mechanical resonators
\cite{Drummond80}-\cite{Chan05}. All these systems display
bistability of forced vibrations. Because of thermal or externally
applied noise, there occurs switching between coexisting stable
vibrational states. The measured switching probabilities of
noise-induced transitions are in a good agreement with the theoretical
predictions.

In the present paper we are interested in the dynamics of a quantum
oscillator. The generality of the oscillator as a model system and the
current interest in quantum computing and coherent phenomena lead to
two major questions: (i) does a resonantly driven oscillator display
coherent quantum effects that would qualitatively differ from those in
two-level systems, and (ii) in the presence of relaxation, what is the
probability of switching between coexisting stable states due to
quantum fluctuations? These two questions are addressed in the present
paper, which is based on the results \cite{DS88}-\cite{Marthaler05}.

A weakly nonlinear oscillator is a multi-level quantum system with nearly equidistant
energy levels $E_n$. Therefore a periodic force of frequency $\omega_F$ can be nearly
resonant for many transitions at a time, i.e., $\hbar\omega_F$ can be close to the
interlevel distance $E_{n+1}-E_n$ for many $n$. This leads to strong nonlinearity of the
response even to comparatively weak resonant fields. A well-known quantum effect of the
oscillator nonlinearity is the onset of Rabi oscillations due to resonant multiphoton
transitions. Multiphoton Rabi oscillations occur when the spacing between remote energy
levels $n$ and $m$ coincides with the energy of $n-m$ photons, $E_n-E_{m} =
(n-m)\hbar\omega_F$ \cite{Bloembergen76}.  The multiphoton transition amplitude is
resonantly enhanced, because an $m\to n$ transition occurs via a sequence of virtual
field-induced transitions $k\to k+1$ (with $m\leq k \leq n-1$), all of which are almost
resonant. An associated classical effect, in the presence of dissipation, is hysteresis
of the amplitude of forced vibrations as function of the field amplitude $A$ and
$\omega_F$.

In this paper (see also \cite{Dykman_Fistul05}) we show that
multiphoton transitions in the oscillator are accompanied by a new
effect, {\it antiresonance} of the response.  When the frequency of
the driving field adiabatically passes through a resonant value, the
vibration amplitude displays a sharp minimum or maximum, depending on
the initially occupied state. We argue that the antiresonance and the
multiphoton Rabi oscillations can be observed in Josephson junctions.

If the frequency $\omega_F$ is close to twice the oscillator
frequency, then $\hbar\omega_F$ is close to $E_{n+2}-E_n$ for many $n$
at a time. This leads to parametric resonance in the oscillator, in
which it oscillates at frequency $\omega_F/2$ in response to the
driving. Such oscillations are intrinsically bistable, because their
phase can take on two values that differ by $\pi$.

We will be most interested in the semiclassical behavior of the
oscillator, which, on the one hand, stretches all the way to the
classical region, and on the other hand, works well for
oscillators even deep in the quantum domain. In the semiclassical
picture, resonant multiphoton transitions correspond to tunneling
between Floquet states of the oscillator with equal quasienergies.
[The quasienergy $\varepsilon$ gives the change of the wave
function $\psi(t)$ when time is incremented by the modulation
period $\tau_F$,
$\psi(t+\tau_F)=\exp(-i\varepsilon\tau_F/\hbar)\psi(t)$]. The
occurrence of equal-quasienergy states is related to the
bistability of forced vibrations of a classical oscillator.

Tunneling of
a driven oscillator is a carefully studied\cite{Dyakonov86} example of
dynamical tunneling \cite{Heller81}. The WKB analysis
gives an important insight into the origin of the antiresonance, which
goes beyond the perturbation theory in the driving field.

Dynamical tunneling also leads to transitions between coexisting metastable states of
forced vibrations, which emerge in the presence of dissipation due to coupling to a
thermal reservoir. In terms of quantum mechanics, dissipation is due to interlevel
oscillator transitions with energy being transferred to (or from, for nonzero
temperature) the reservoir. It turns out that dissipation may also lead to transitions
between metastable states of forced vibrations, even for zero temperature \cite{DS88}.

For $T=0$ there occur only interlevel transitions where the oscillator
energy goes to the reservoir (but the energy loss is compensated by
the driving field, in the stationary regime). However, the
quasi-energy may increase or decrease as a result of a
coupling-to-reservoir induced transition, although with different
probabilities.  Therefore along with drift over quasienergy towards a
metastable state, which results from more probable transition, there
emerges diffusion away from this state as a sequence of less probable
transitions. The diffusion may lead to activated-like
escape. Activation in this case has purely quantum nature, and
therefore we call it quantum activation.

\section{The Models}

The Hamiltonian of a nonlinear oscillator with mass $M=1$ has the form
\begin{equation}
\label{Hamiltonian_full} H(t)=\frac{1}{2}p^2 + \frac{1}{2}\omega_0^2q^2 +
\frac{1}{4}\gamma q^4 + H_F(t).
\end{equation}
We will consider two types of periodic modulation, $H_F=H_F^{(r,p)}$,
which correspond to resonant and parametric driving,
\begin{eqnarray}
\label{H_F} H_F^{(r)}(t)= -qA\cos(\omega_Ft), \qquad
\delta\omega^{(r)} \equiv \omega_F-\omega_0\ll \omega_0,\\
H_F^{(p)}=\frac{1}{2}q^2F\cos(\omega_Ft),\qquad
\delta\omega^{(p)} \equiv \frac{1}{2}\omega_F-\omega_0\ll
\omega_0\nonumber
\end{eqnarray}
(in what follows we set $\gamma > 0$).

It is convenient to analyze the dynamics in the rotating wave
approximation by switching from the fast oscillating operators $q,p$
to slowly varying operators $Q,P$ using transformations
\[q=\alpha^{(r)}(Q\cos\omega^{(r)} t+P\sin\omega^{(r)} t), \qquad
p=-\alpha^{(r)}\omega^{(r)}
(Q\sin\omega^{(r)} t - P\cos\omega^{(r)} t) \]
for resonant driving and
\[q=\alpha^{(p)}(P\cos\omega^{(p)} t - Q\sin\omega^{(p)} t), \qquad
p=-\alpha^{(p)}\omega^{(p)}
(P\sin\omega^{(p)} t + Q\cos\omega^{(p)} t) \]
for parametric driving, with
\[\omega^{(r)} =
\omega_F, \qquad \alpha^{(r)}=(8\omega_F\delta\omega^{(r)}/3\gamma)^{1/2},\]
and
\[\omega^{(p)} = \omega_F/2, \qquad
\alpha^{(p)}=(2F/3\gamma)^{1/2}.\]

The variables $Q,P$ are the appropriately scaled coordinate and
momentum. The commutation relation
for them has a simple form
\begin{equation}
\label{commutator}
[P,Q]=-i\lambda,\qquad \lambda =\hbar\left(\omega^{(r,p)}\right)^{-1}
\left(\alpha^{(r,p)}\right)^{-2}.
\end{equation}
The parameter $\lambda$ plays the role of the effective Planck
constant. We note that it is proportional to the oscillator
nonlinearity $\gamma$ scaled by the comparatively small detuning of
the field frequency, in the case of nearly resonant driving, or the
comparatively small field amplitude, in the case of parametric
driving.

\begin{figure}[h]
\includegraphics[width=3.0in]{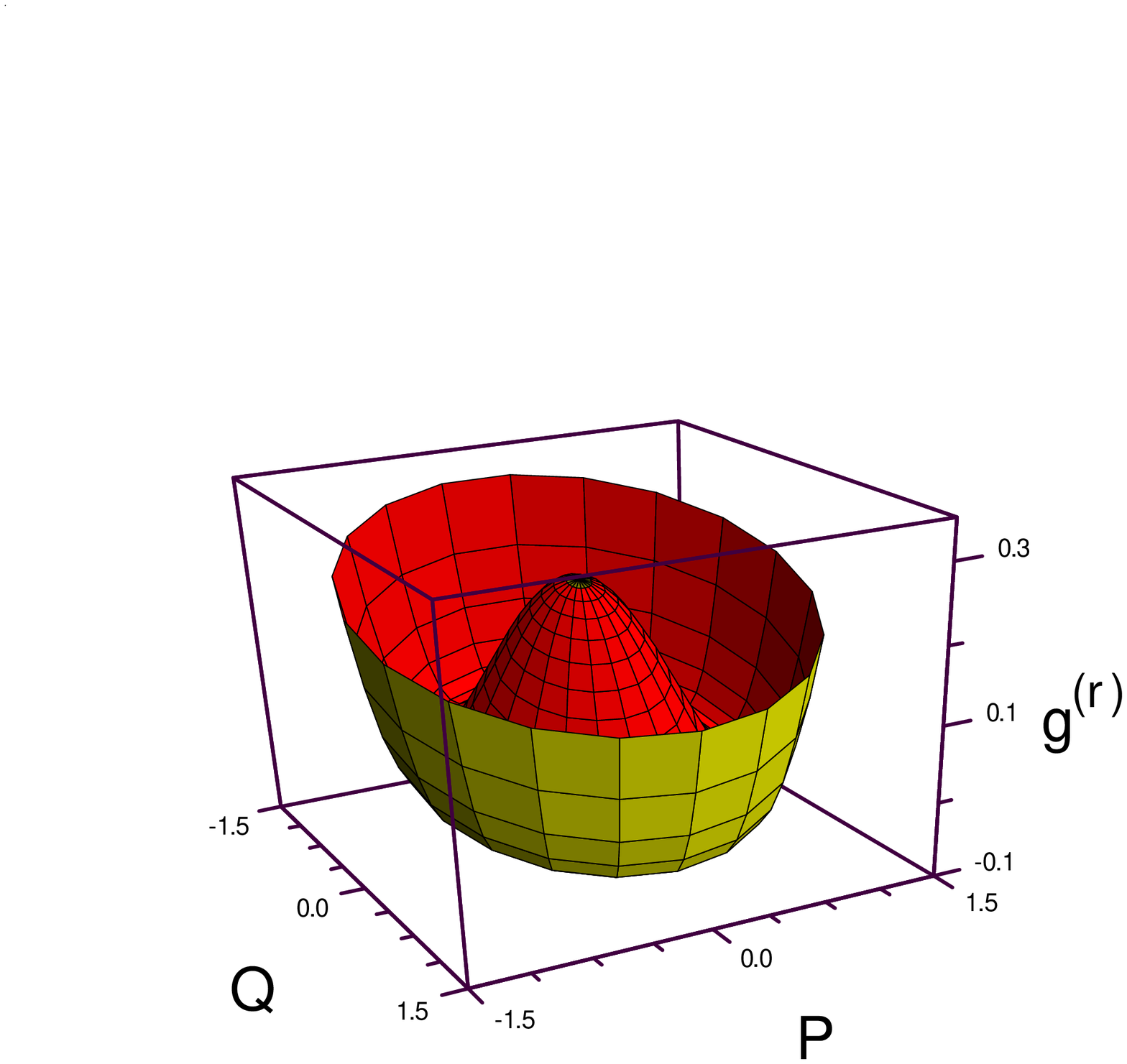}
\includegraphics[width=3.0in]{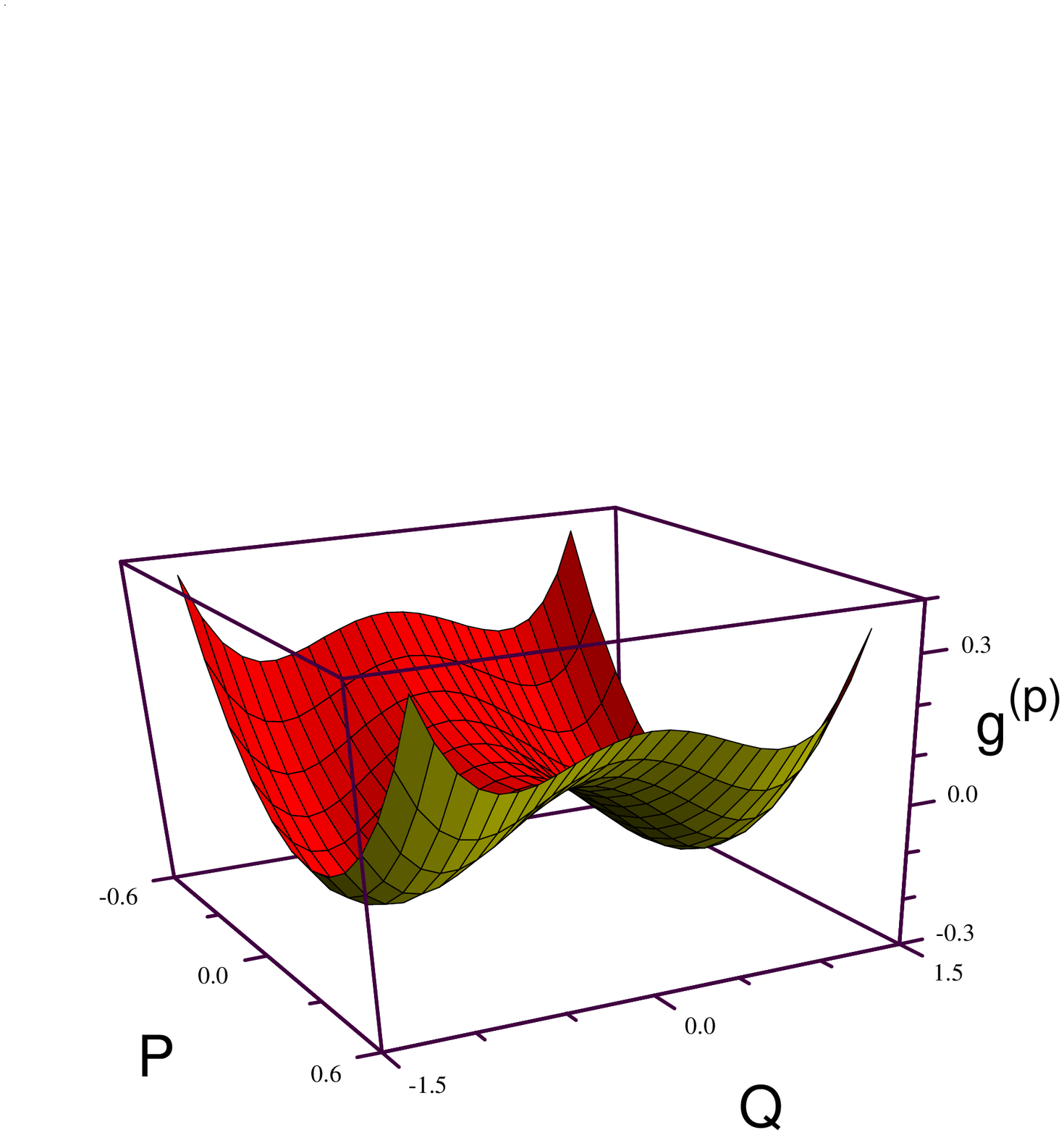}
\caption{The scaled classical quasienergy of the oscillator for
resonant and parametric driving (left and right panels, respectively)
as function of the slowly varying coordinate and momentum $Q,P$. The
plot of $g^{(r)}$ refers to the reduced field intensity $\beta=
3\gamma A^2/32\omega_F^3(\delta\omega^{(r)})^3 = 1/27$. The plot of
$g^{(p)}$ refers to $\mu= 2\omega_F\delta\omega^{(p)}/F=-0.1$. }
\label{fig:g_function}
\end{figure}

The dynamics of $Q,P$ in the two cases is described by effective
Hamiltonians
\begin{eqnarray}
\label{eff_Hamiltonians}
H^{(r)}=
\omega_F\delta\omega^{(r)}\left(\alpha^{(r)}\right)^{2}g^{(r)},\nonumber\\
H^{(p)}=(F/4)\left(\alpha^{(p)}\right)^{2}g^{(p)}.
\end{eqnarray}
Their eigenvalues are equal to the  quasienergies
$\varepsilon_n$ of the oscillator. The functions $g^{(r,p)}$ have the
forms
\begin{eqnarray}
g^{(r)}(P,Q)=\frac{1}{4}(Q^2+P^2-1)^2 - \beta^{1/2}Q, \quad
\beta=\frac{3\gamma A^2}{32\omega_F^3\left(\delta\omega^{(r)}\right)^3},\\
g^{(p)}=\frac{1}{4}(Q^2+P^2)^2+ \frac{1}{2}(1-\mu)P^2 -
\frac{1}{2}(1+\mu)Q^2,\quad
\mu=2\omega_F\delta\omega^{(p)}/F.
\end{eqnarray}
They are shown in Fig.~\ref{fig:g_function}. Each of them depends on one parameter. In
the region of bistability of period one vibrations, $0 < \beta <4/27$, the function
$g^{(r)}$ has a shape of a tilted Mexican hat, with a maximum at the top of the central
dome and a minimum at the lowest point of the rim. For a parametrically excited
oscillator in the region $-1 < \mu < 1$ the function $g^{(p)}$ has two symmetrical
minima. These extrema of $g^{(r,p)}$ correspond to metastable states of the driven
oscillator in the presence of weak dissipation. The saddle points of $g^{(r,p)}$
correspond to unstable states of forced vibrations.

\section{Multiphoton antiresonance}

We will start with the studies of the coherent response
of the oscillator to a nearly resonant field.
When the field amplitude $A\to 0$, the eigenstates $|n\rangle$ of the
Hamiltonian $H^{(r)}$ coincide with the Fock states of the oscillator,
and the quasienergies are
\begin{equation}
\label{varep_A_0}
\ep_n = -\delta\omega^{(r)}n+ \frac{1}{2}Vn(n+1), \quad
V=\frac{3\hbar\gamma}{4\omega_0^2}.
\end{equation}
We keep only the lowest-order term in $V$, which corresponds to the
weak nonlinearity approximation. In this approximation the energy of
an $N$th oscillator state for $A=0$ is $E_N=\hbar\omega_0N +
VN(N+1)/2$.  The $N$-photon resonance $N\hbar\omega_F=E_N-E_0$ occurs,
in terms of $\dr$, for
\[\dr= \dr_N =V(N+1)/2.\]
For the corresponding field frequency $\omega_F$ the quasienergies
$\ep_0$ and $\ep_N$ are equal.

\begin{figure}[h]
\includegraphics[width=2.7in]{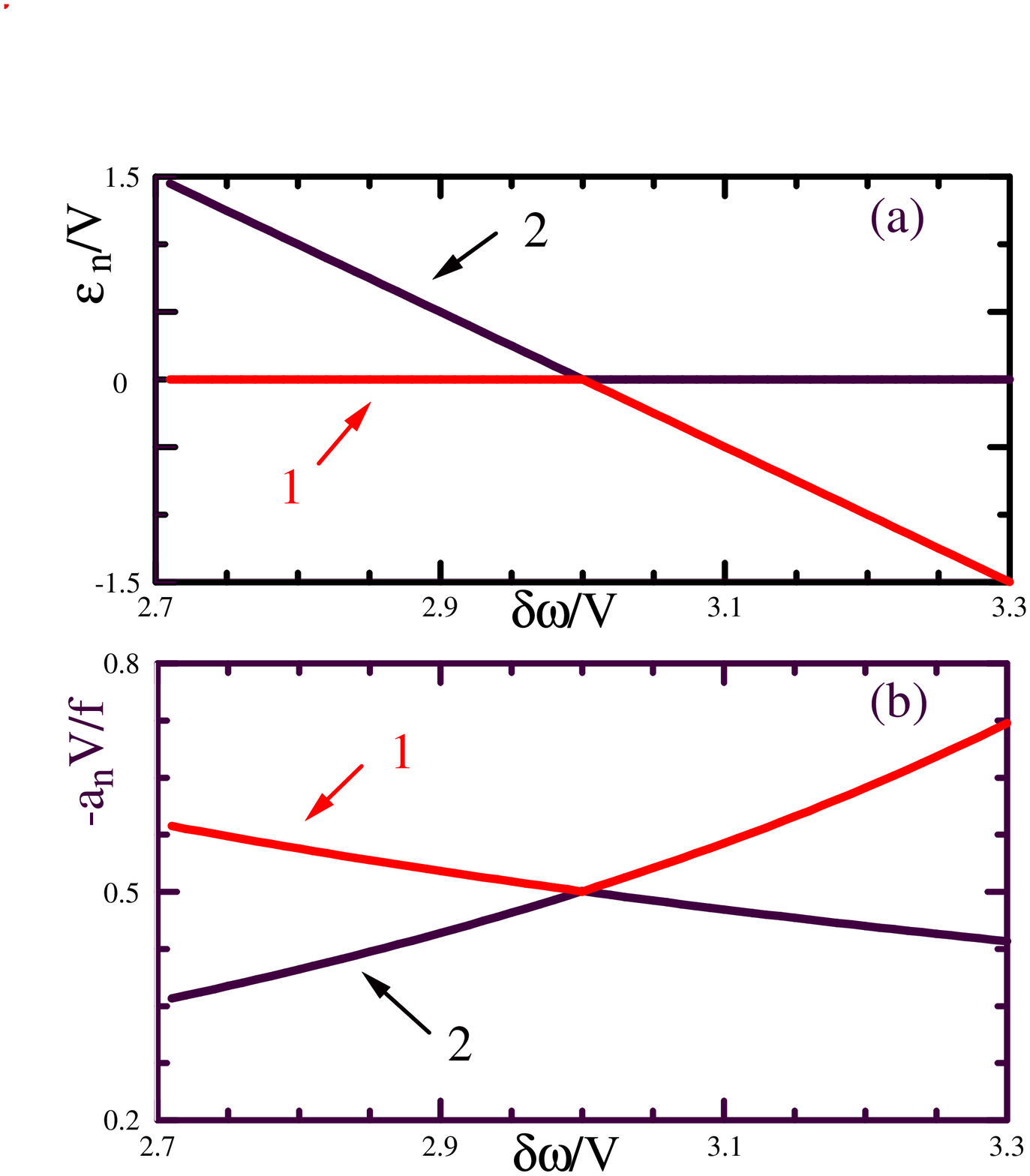}
\includegraphics[width=3.0in]{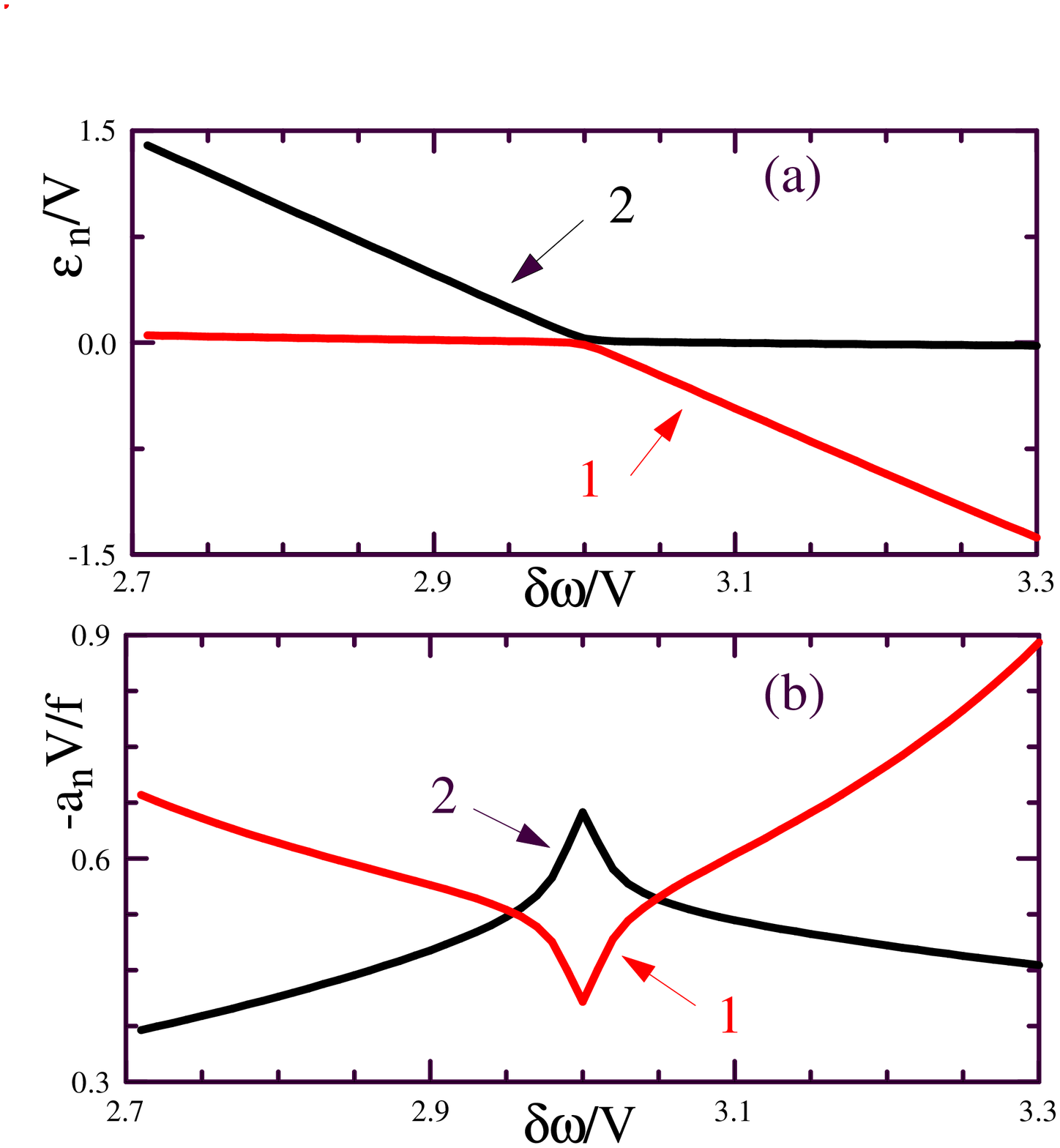}
\caption{Quasienergy levels $\ep_n$ and reduced susceptibilities in the resonating states
of a driven nonlinear oscillator. The plots refer to a 5-photon resonance, $N=5$. The
labels 1 and 2 correspond to the ground ($n=0$) and 5th ($n=5$) oscillator states in the
absence of driving, for $\delta\omega \equiv \dr < \dr_5$. Left panel: the limit of zero
driving amplitude, $A\to 0$. The quasienergy levels cross for $\dr_N/V = 3$. The
susceptibilities in the resonating states also cross at that same frequency. Right panel:
level repulsion and antiresonance of susceptibilities for comparatively weak modulation.
The data refer to $A/A_N=1/2$. It is seen from the figure that the susceptibility
antiresonance is much more pronounced than the level repulsion. The scaling factor for
the amplitude $a_n$ of oscillator vibrations in the states $n=0,5$ is
$f=(8\hbar\omega_0)^{-1/2}A$.} \label{fig:en_susc}
\end{figure}

The field leads to mixing of the wave functions of resonating states
and to level anticrossing. This anticrossing is clearly seen in the
upper right panel of Fig.~\ref{fig:en_susc}. The minimal splitting of
the levels $\varepsilon_0$ and $\varepsilon_N$ is given by the
multiphoton Rabi frequency $\Omega_R$. For weak field it can be
obtained by perturbation theory \cite{Bloembergen76}. To the lowest
order in the field amplitude $A$ in the limit of large $N$
\begin{eqnarray}
\label{Rabi}
\Omega_R=V\,(A/A_N)^N N^{5/4}(2\pi)^{-3/4},\qquad
A_N=(2\hbar\omega_0)^{1/2}|V|\,N^{3/2}\exp(-3/2)/2.
\end{eqnarray}
The Rabi frequency depends on $N$ exponentially, $\Omega_R\propto
A^N$. This dependence works well numerically in the whole region $A<
A_N$ \cite{Dykman_Fistul05}.

Coherent response of the oscillator to the driving field is
characterized by the expectation value of the coordinate $q$. If the
oscillator is in an eigenstate $|n\rangle$ of the Hamiltonian
$H^{(r)}$, this value has the form
\begin{equation}
\label{a_n}
q_n=(\hbar/2\omega_0)^{1/2}a_ne^{-i\omega_Ft}+{\rm c.c.},
\qquad q_n=\langle n|q|n\rangle.
\end{equation}

To first order in the field, the reduced amplitude of forced
vibrations in an $n$th state $a_n$ is
\begin{equation}
\label{a_first_order}
a_n=-f\dr/ \bigl((\dr
-Vn)[\dr-V(n+1)]\bigr),  \qquad f=(8\hbar\omega_0)^{-1/2}A.
\end{equation}
Remarkably, for $\dr =\dr_N$ the vibration
amplitudes in the resonating states coincide with each other,
$a_{N-n}=a_n$ for $0\leq n <N/2$, see the left lower panel in
Fig.~\ref{fig:en_susc}.

Field-induced multiphoton mixing leads not only to splitting of the
quasienergy levels, but also to repulsion of the vibration
amplitudes. It can be calculated by diagonalizing the Hamiltonian
$H^{(r)}$ and is shown in the right lower panel of
Fig.~\ref{fig:en_susc} as a function of frequency detuning
$\delta\omega\equiv \dr$. One of the involved resonating states is the
ground state of the oscillator $n=0$ in the limit $A\to 0$. The
quantities plotted in Figs.~\ref{fig:en_susc}(b) are susceptibilities,
they are proportional to the ratio of the vibration amplitude to the
modulation amplitude $a_n/A$.

The antiresonant splitting of the expectation values of the vibration
amplitudes is by far the most interesting feature of
Fig.~\ref{fig:en_susc}. It occurs at the adiabatic passage of $\dr$
through resonance, where the system switches between the ground and
excited states. In particular, the amplitude displays an antiresonant
dip if the oscillator is mostly in the ground state for $(\dr-\dr_N)/V
<1$ or in the state $N$ for $(\dr-\dr_N)/V >1$. The magnitude and
sharpness of the dip are determined by $\Omega_R/V$ and depend very
strongly on the field and $N$. With decreasing $\Omega_R/V$ the dip
(and peak) start looking like cusps located at resonant frequency. We
note that, in contrast to the case of energy levels, there is no
reason for repulsion (anticrossing) of susceptibilities. In fact, as
seen from Fig.~\ref{fig:en_susc} the susceptibilities do cross,
although away from the resonant frequency. The effect of antiresonance
is due purely to specific quantum interference \cite{Dykman_Fistul05}.

The dip in the oscillator response has no analogue in two-level
systems. There, for nearly resonant driving, the coherent response in
the two adiabatic states differs only in sign. It displays a
peak when the radiation frequency adiabatically passes through the
transition frequency.

\subsection{The WKB picture of the antiresonance}

In the WKB approximation, Rabi oscillations correspond to tunneling between the states
with nearly equal quasienergies. Such semiclassical states can be found from the
Hamiltonian $H^{(r)}$ (\ref{eff_Hamiltonians}) by using the Bohr-Sommerfeld quantization
condition applied to the mechanical action $\oint P\,dQ$ for trajectories
$g(Q,P)=$~const, with $\hbar$ replaced by $\lambda$, Eq.~(\ref{commutator}). It is seen
from the left panel of Fig.~\ref{fig:g_function} that, in a certain range of $g$, there
are two types of trajectories with the same $g$, those on the internal dome and those on
the external part of the Mexican hat $g(Q,P)$. If, as a result of the Bohr-Sommerfeld
quantization, the quantized values of $g$ on the two parts of the surface $g(Q,P)$
coincide with each other, then there may occur resonant  tunneling between the
corresponding quantum states. The resulting tunneling splitting \cite{Dyakonov86} is the
Rabi frequency.

Interestingly, one can show that the average value of the coordinate
\[Q(g)=\tau^{-1}(g)\int\nolimits_0^{\tau(g)}Q(t)\,dt\]
[$\tau(g)$ is the period of oscillations for a given $g$] is the same
for the internal and external trajectories with the same $g$. This
corresponds to the susceptibilities of the resonating quasienergy
states being the same, in the neglect of tunneling-induced mixing of
the states. We emphasize that the fact that the susceptibilities are
equal is not a result of the perturbation theory in the field
amplitude, as in the case of Eq.~(\ref{a_first_order}), they are equal
in all orders of the perturbation theory in $A$ as long as tunneling
is disregarded. Tunneling-induced state mixing leads to the
antiresonance of the response \cite{Dykman_Fistul05}.

\section{Escape of a driven system: tunneling or quantum activation?}

We will now briefly outline the new effects and unanswered questions
that emerge when dissipation is taken into account
\cite{DS88,Marthaler05}. We will assume that, even though dissipation
is weak, the dissipation rate exceeds the tunneling rate. The problem
of fairly general interest that will be addressed is switching between
metastable states of forced vibrations of a quantum oscillator. We
will consider the most interesting situation where there are many
quasienergy states between the extrema of the quasienergy surface. In
the case of escape of a particle from a potential well it corresponds
to a well with many energy levels.

\begin{figure}[h]
\includegraphics[width=1.0in]{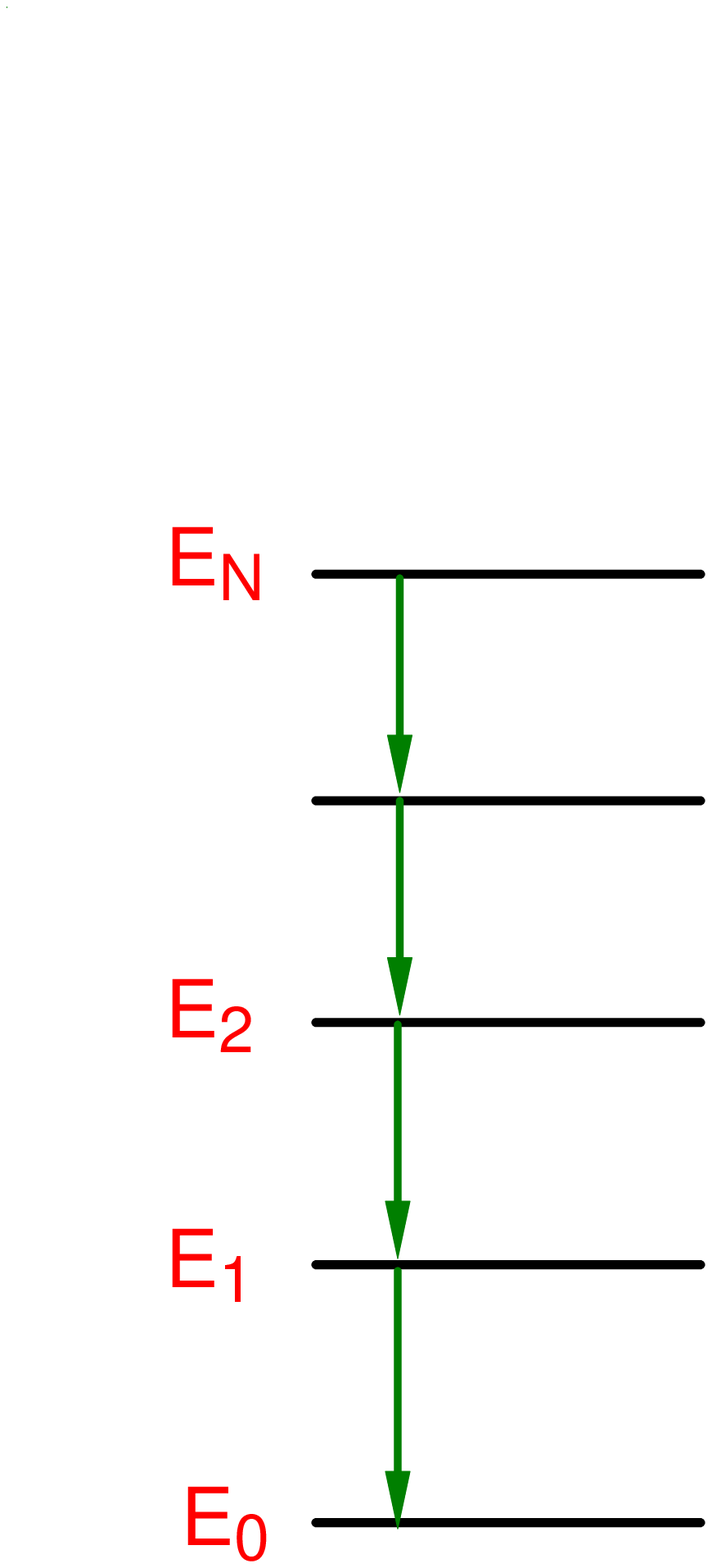}\hspace*{0.2in}
\includegraphics[width=1.9in]{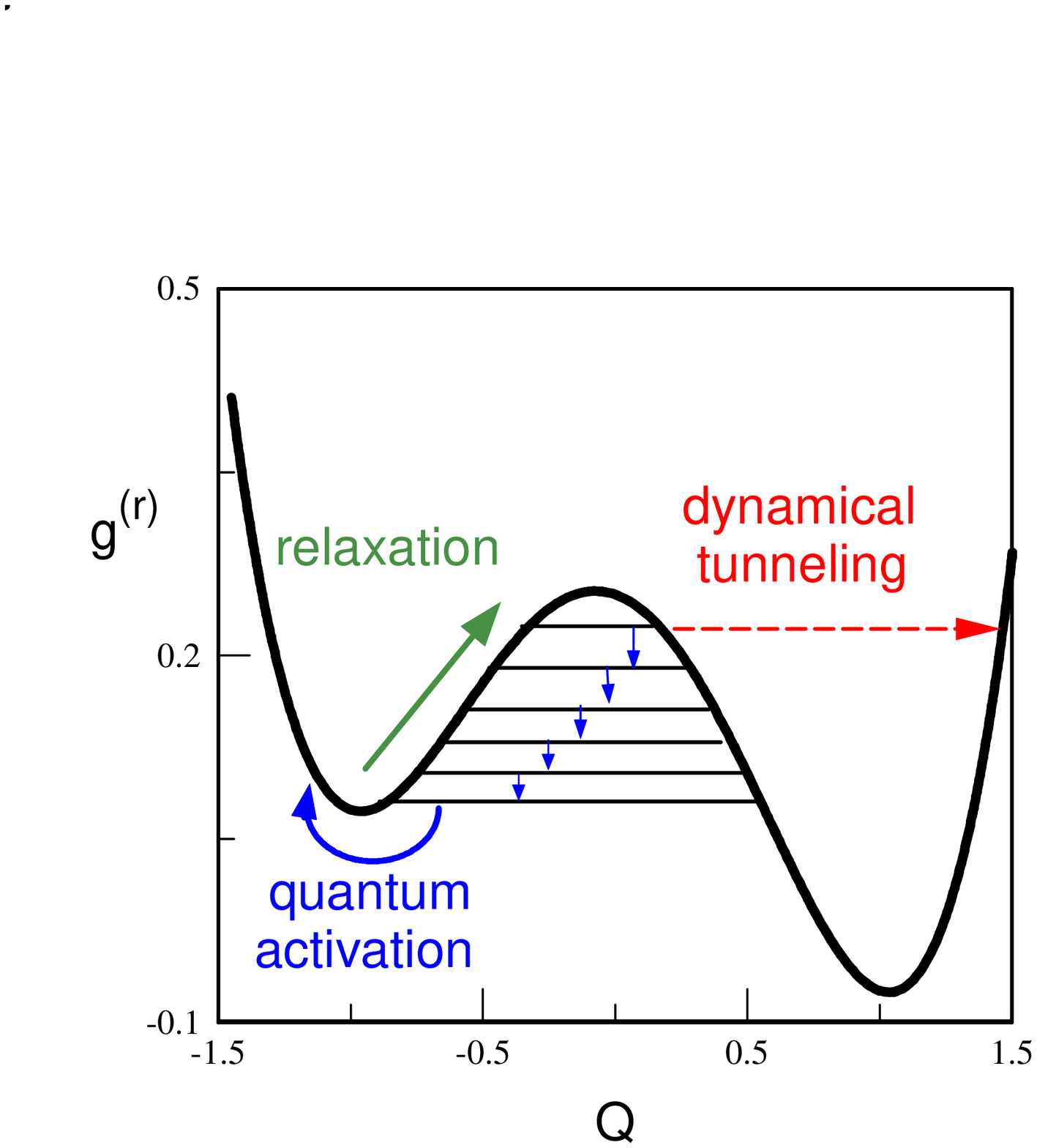}\hspace{0.2in}
\includegraphics[width=2.2in]{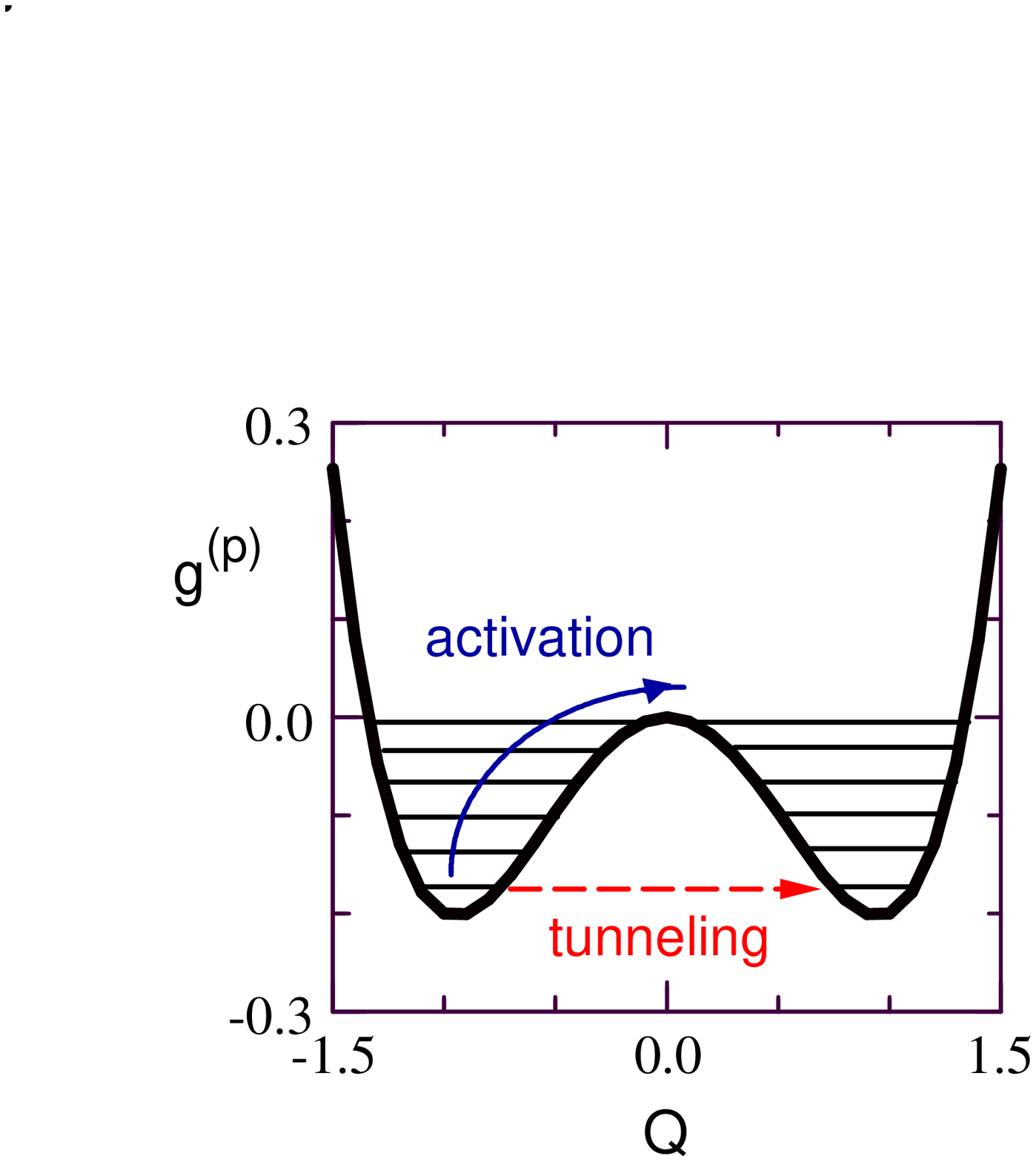}
\caption{Left panel: transitions between the energy levels of the oscillator $E_n$ due to
coupling to a thermal reservoir. For $T=0$ there occur only transitions to lower levels
with emission of excitations of the reservoir. In the simplest case relaxation is due to
transitions between neighboring oscillator levels. Central panel: the cross-section of
the surface of the scaled quasienergy $g^{(r)}$ shown in
Fig.~\protect\ref{fig:g_function} by the plane $P=0$. Quasienergy levels for orbits that
lie on the dome in Fig.~\protect\ref{fig:g_function} are shown schematically. Relaxation
to the metastable state at the top of the dome corresponds to transitions to higher
quasienergy, whereas less probable transitions to lower quasienergy lead to an
activation-type escape from the metastable state. Right panel: the cross-section of the
quasienergy surface of the parametrically driven oscillator $g^{(p)}$,
Fig.~\protect\ref{fig:g_function}, by the plane $P=0$, with schematically shown
quasienergy levels. Relaxation drives the oscillator down to the lowest quasienergy
levels at the bottom of the wells of $g^{(p)}(Q,P)$, but there is a possibility of
activation even for $T=0$.} \label{fig:tun_escape}
\end{figure}

In systems in thermal equilibrium, the rate of tunneling decay of a
metastable state for low temperatures is given by the probability of a
tunneling transition from the ground state in a metastable potential
well. In the case of a resonantly driven oscillator this corresponds
to dynamical tunneling from the top of the dome of the quasienergy
surface $g^{(r)}$ to the state on the external orbit with the same
quasienergy, see the left panel of Fig.~\ref{fig:g_function}. The
tunneling is shown schematically in the central panel of
Fig.~\ref{fig:tun_escape}. For a parametrically driven oscillator the
corresponding tunneling occurs between the minima of the surface
$g^{(p)}$, as shown in the right panel of Fig.~\ref{fig:tun_escape}.

For higher temperatures, again in the case of equilibrium systems, one has to take into
account tunneling from excited intrawell states. Escape may occur also via thermal
activation over the potential barrier. One of these escape mechanisms dominates,
depending on temperature \cite{Larkin85}. In the case of a driven oscillator tunneling
from excited states corresponds to tunneling with quasienergies that differ from those at
the extrema of $g(Q,P)$. In addition, there is a probability of activation over the
quasienergy barrier. However, since the distribution over quasienergy is not of the
Boltzmann form, it is not clear which of the escape mechanisms dominates at a given
temperature of the bath.

For weak dissipation the distribution can be described by the balance
equation for the occupations $\rho_n$ of quasienergy (Floquet) states
$|n\rangle$,
\begin{equation}
\label{balance} \dot \rho_n=-\sum\nolimits_m
W_{nm}\rho_n+\sum\nolimits_m W_{mn}\rho_m.
\end{equation}
The transitions probabilities $W_{nm}$ can be calculated as matrix elements of the
operator that describes relaxation of the oscillator. The wave functions $|n\rangle$ can
be found from the Bohr-Sommerfeld approximation disregarding tunneling. A standard WKB
calculation allows one to express $W_{nm}$ in terms of the Fourier components of the
coordinate and momentum $Q_{m-n}(g_n), P_{m-n}(g_n)$ for a given quasienergy $g_n$ [the
functions $Q(t), P(t)$ are periodic functions of time for a given $g$, with period
$\tau(g)$]. In the simple case of linear friction, which corresponds to relaxation
transitions between nearest energy (not quasienergy) levels of the oscillator shown in
the left panel of Fig.~\ref{fig:tun_escape}, $W_{nm}$ are simply quadratic in
$Q_{m-n}(g_n), P_{m-n}(g_n)$. They exponentially decay with $|n-m|$.

The probabilities $W_{nm}$ are organized so that it is more likely for
a system to make a transition toward the value of $g$ in the
metastable state rather than away from it. This is why the state is
metastable. However, in contrast to systems in equilibrium,
the probabilities $W_{nm}$ do not satisfy the condition
$W_{nm}=W_{mn}\exp[(g_n-g_m)/kT]$.  Even for $T\to 0$ there is a
nonzero probability to make a transition in the direction opposite to
the metastable state. This is a consequence of the fact that the
Floquet states $|n\rangle$ are linear combinations of the Fock states
of the oscillator. Therefore, even where all transitions between the
Fock states go in one direction in energy, as in the left panel of
Fig.~\ref{fig:tun_escape}, transitions between the Floquet states go
in different directions in quasienergy, although with different
probabilities.

Transitions in the ``wrong'' direction lead to diffusion over
quasienergy away from the metastable state. Their immediate
consequence is a finite width of the stationary distribution over
quasienergy even for $T\to 0$. Another closely related consequence is
a nonzero probability to reach quasienergy of the saddle of the
functions $g^{(r,p)}$ starting from a metastable state. The saddle of
$g^{(r,p)}$ is similar to the top of a potential barrier for a
particle in a metastable potential. The logarithm of the probability
of reaching the saddle gives the activation energy of escape as a
result of diffusion over quasienergy. We call this process quantum
activation, since it occurs even for $T=0$.

For resonantly and parametrically driven oscillators the transition rates $W_{nm}$ can be
calculated explicitly using the fact that the classical trajectories in $(Q,P)$-variables
can be expressed in terms of the Jacobi elliptic functions \cite{DS88,Marthaler05}. We
have compared the activation energies of escape with the tunneling exponents. When the
activation energy is smaller than the tunneling exponent in the absolute value, escape
occurs via quasienergy diffusion. We found that, both for a resonantly and a
parametrically excited oscillator, escape from metastable states occurs via activation,
not tunneling. This holds true for all parameter values where a nonlinear oscillator has
coexisting stable states. We found an unusual behavior of the distribution for an
underdamped parametrically driven oscillator for $T=0$. For some quasienergy the
distribution displays a sharp decrease, and at the same time the tunneling rate goes to
zero.

The physical origin of the fact that escape occurs via activation, not
dynamical tunneling, remains not fully understood, the existing
arguments are formal \cite{DS88}. Apparently, there must be a
crossover from activation to tunneling when the system goes to
equilibrium, but the models that we have discussed are strongly
nonequilibrium, the very presence of coexisting metastable states is
due to periodic driving.

In conclusion, we have shown that a simple system, a driven nonlinear oscillator,
displays unusual quantum coherent phenomena and unusual switching behavior. The
considered effects have no analogue in two-level systems and are qualitatively different
from what has been known about switching in thermal equilibrium systems. They are not
only of fundamental interest, but are also important for many applications, in particular
in sensing and quantum measurements.

\begin{theacknowledgments}
  This research was supported in part by the Institute for Quantum
Sciences at Michigan State University and by the NSF through grant
ITR-0085922.
\end{theacknowledgments}


\begin{thebibliography}{99}

\bibitem{Drummond80} P. D. Drummond and D. F. Walls, \emph{J. Phys. A} {\bf
13}, 725 (1980).

\bibitem{Gabrielse85} G. Gabrielse, H. Dehmelt, and W. Kells,
\emph{Phys. Rev. Lett.} {\bf 54}, 537 (1985).

\bibitem{Larkin_all} A.I. Larkin and Yu.N.~Ovchinnikov, \emph{J. Low
Temp. Phys.}  {\bf 63}, 317 (1986); B.I.~Ivlev and V.I.~Mel'nikov,
\emph{Phys. Lett. A} {\bf 116}, 427 (1986); S.~Linkwitz and H.~Grabert,
\emph{Phys. Rev. B} {\bf 44} 11888, 11901 (1991);
M.H. Devoret, D. Esteve, J.M. Martinis, A. Cleland, and J. Clarke,
\emph{Phys. Rev. B} {\bf 36}, 58-73 (1987).

\bibitem{JosEscape} A. Wallraff, T. Duty, A. Lukashenko, and A. V. Ustinov,
\emph{Phys. Rev. Lett.} \textbf{90}, 037003
(2003); M. V. Fistul, A. Wallraff, and A. V. Ustinov, \emph{Phys. Rev. B},
{\bf 68}, 060504(R) (2003).

\bibitem{Siddiqi04} I. Siddiqi,
R. Vijay, F. Pierre, C. M. Wilson,
M. Metcalfe, C. Rigetti, L. Frunzio, and M. H. Devoret,
\emph{Phys. Rev. Lett.} {\bf 93}, 207002 (2004).

\bibitem{Cleland05} J. S. Aldridge and A. N. Cleland,
\emph{Phys. Rev. Lett.} {\bf 94}, 156403 (2005); cond-mat/0406528
(2004).

\bibitem{Mohanty_APL05}R. L. Badzey, G. Zolfagharkhani, A. Gaidarzhy,
and P. Mohanty, \emph{Appl. Phys. Lett.} {\bf 86}, 023106 (2005).

\bibitem{Chan05} C. Stambaugh and H. B. Chan, cond-mat/0504791 (2005).

\bibitem{DS88} M.I. Dykman and V.N. Smelyanskiy,
\emph{Zh. Eksp. Teor. Fiz.} {\bf 94}, 61 (1988) [\emph{Sov. Phys. JETP} {\bf 67},
1769 (1988)].

\bibitem{Dykman_Fistul05} M.I. Dykman and M.V. Fistul, \emph{Phys. Rev. B} {\bf 71}, 140508(R) (2005).

\bibitem{Marthaler05} M.I. Dykman and M. Marthaler, unpublished.

\bibitem{Bloembergen76} D.M. Larsen and N. Bloembergen,
\emph{Opt. Communications} {\bf 17}, 254 (1976).

\bibitem{Dyakonov86} A.P. Dmitriev and M.I. D'yakonov,
\emph{Zh. Eks. Teor. Fiz.} {\bf 90}, 1430 (1986)
[\emph{Sov. Phys. JETP} {\bf 63}, 838 (1986)].

\bibitem{Heller81} M. J. Davis and E. J. Heller, \emph{J. Chem Phys.} {\bf
75}, 246 (1981).

\bibitem{Larkin85} A. I. Larkin and Y. N. Ovchinnikov, J. Stat. Phys. {\bf 41}, 425 (1985)


\end{thebibliography}
\end{document}